%% file: BAD-2317.tex
\documentclass[twocolumn,showpacs,aps,prl,superscriptaddress]{revtex4}

\usepackage{graphicx}
\usepackage{dcolumn}
\usepackage{epsfig}
\usepackage{psfrag}
\usepackage{amsmath}

\newcommand{\BaBarYear}    {10}
\newcommand{\BaBarNumber}  {029}

\newcommand{\SLACPubNumber} {14342}

 \newcommand{\BaBarType}      {PUB}  

\input pubboard/babarsym   

\input Q2BDefn/Definitions.tex

\newcommand{\bsg}{\ensuremath{b\rightarrow s g}}

\newcommand{\BPiX}{\ensuremath{B\rightarrow\pi^+ X}}

\newcommand{\BKpX}{\ensuremath{B\rightarrow K^+ X}}

\newcommand{\BKzX}{\ensuremath{B\rightarrow K^{0} X}}

\newcommand{\pstar}{\ensuremath{p^{*}}}

\newcommand{\bsig}{\ensuremath{B_{\rm sig}}}
\newcommand{\breco}{\ensuremath{B_{\rm reco}}}

\newcommand{\pBRBKpX}{\ensuremath{119^{+32}_{-29} \pm 37}}        
\newcommand{\ulpBRBKpX}{\ensuremath{187}}                    
\newcommand{\signifBKpX}{\ensuremath{2.9}}                     
\newcommand{\AcpBKpX}{\ensuremath{0.57\pm0.24\pm0.05}}       

\newcommand{\pBRBKsX}{\ensuremath{195^{+51}_{-45} \pm 50}}        
\newcommand{\ulpBRBKsX}{\ensuremath{294}}                    
\newcommand{\signifBKsX}{\ensuremath{3.8}}                     

\newcommand{\pBRBPiXn}{\ensuremath{372^{+50}_{-47} \pm 59 }} 
\newcommand{\signifBPiX}{\ensuremath{6.7}}                     
\newcommand{\AcpBPiX}{\ensuremath{0.10\pm0.16\pm0.05}}       

\newcommand{\theTitle}{{\boldmath Measurement of partial branching fractions of
inclusive charmless \B\ meson decays to \Kp, \Kz, and \pip}}

\begin{document}


\begin{flushleft}
\babar-\BaBarType-\BaBarYear/\BaBarNumber \\
SLAC-PUB-\SLACPubNumber \\
\end{flushleft}

\title{\theTitle}

\input pubboard/authors_oct2010_bad2317.tex

\date{\today}

\begin{abstract}
We present measurements of partial branching fractions of
\BKpX, \BKzX, and \BPiX, where $X$ denotes any accessible final
state above the endpoint for $B$ decays to charmed mesons,
specifically for momenta of the candidate hadron greater than 2.34 
(2.36) GeV for kaons (pions) in the \B\ rest frame.
These measurements are sensitive to potential new-physics 
particles which could enter the $b \ra s(d)$ loop transitions. 
The analysis is performed on a data sample consisting of 383 
\timesix\ \BB\ pairs collected with the \babar\ detector at the 
PEP-II \epem\ asymmetric energy collider. Our results are in 
agreement with standard model predictions and exclude large 
enhancements of the inclusive branching fraction due to sources 
of new physics. 
\end{abstract}

\pacs{13.25.Hw, 12.15.Ji, 11.30.Er}

\maketitle


$B$ mesons decay predominantly to charmed mesons through the tree
level process $b \ra c$, while the tree amplitude $b \ra u$ and 
the one-loop processes $b \ra s$ and $b \ra d$ are strongly 
suppressed. In the standard model (SM), the inclusive 
branching fraction of $B$ mesons to charmless final states is of 
the order of 2\% \cite{greub}.  Particles associated with physics
beyond the SM, such as supersymmetric partners of SM particles, 
could enter the loop amplitudes while leaving the tree-level processes
nearly unaffected, making a sizable enhancement of the 
inclusive $b \ra s\, (d)\; g$ (where $g$ denotes a gluon) branching
fraction possible \cite{bigi,goksu}. Additionally, since
semi-inclusive processes are usually affected by smaller hadronic 
uncertainties than those that arise in calculations
for exclusive final states, these decays can be sensitive to 
nonperturbative amplitudes, such as charming penguins \cite{chay}.

An interesting theoretical mechanism that can modify the SM 
prediction is provided by the Randall-Sundrum
framework, in particular from the Warped Top-Condensation Model
where a radion field $\phi$ is postulated. In the case where 
$1 < m(\phi) < 3.7$ GeV, the radion would decay dominantly to 
gluons, thus enhancing the rate of the charmless $B$ decays through 
the process $b \ra s \phi$. In such a model the 
$b \ra s$ inclusive decay rate could be enhanced by an order of 
magnitude with respect to the SM predictions \cite{radion}.

Historically, an enhancement of charmless $B$ decays had been
postulated \cite{lenz} to explain the deficit of $b \ra c$
processes observed by the ARGUS and CLEO experiments 
\cite{charm_deficit}. Later measurements and refined theoretical
calculations established that no significant discrepancy was 
present \cite{czarnecki}. Inclusive \bsg\ decays have been searched for
by the ARGUS, CLEO, and DELPHI collaborations \cite{prev_meas}.
None of these experiments has found a statistically significant
signal and only upper limits in agreement with theoretical
expectations were set.

In this paper we present measurements of partial branching fractions 
of inclusive charmless \B-meson decays. The signature of these decays 
is the presence of a light meson (\Kp, \KS, or \pip\ \cite{CC}) with momentum 
beyond the kinematic endpoint for \B decays to charmed mesons, measured 
recoiling against a fully reconstructed \B meson. It is possible 
to compare our results with the inclusive branching fraction of $b \ra 
s \gamma$ in the same kinematical region and with some recent theoretical 
predictions \cite{chay} based on Soft Collinear Effective Theory.

The measurement is performed on a data sample collected by
the \babar\ detector \cite{BABARNIM}, operated at the asymmetric 
energy \epem\ PEP-II collider at the SLAC National Accelerator Laboratory. 
We use 347 fb$^{-1}$ (equivalent to 383 \timesix\ \BB\ pairs) collected 
at a center-of-mass energy $\sqrt{s}$ corresponding to the mass of the
\FourS\ resonance, which predominantly decays to charged or
neutral \BB\ pairs; a smaller sample (37 fb$^{-1}$) of data 
collected at an energy of 40 MeV below the \FourS\ peak is used
to study the background originating from continuum
$\epem \ra \qqbar$ ($q = u, d, s, c$) processes.

In order to achieve the highest possible control over the 
backgrounds (mostly arising from continuum), we fully reconstruct 
one of the two $B$ mesons (denoted by \breco) and search for a high 
momentum light hadron (\Kp, \KS, or \pip) among the decay products 
of the other $B$ (\bsig). The full reconstruction of the \breco\
candidate allows us to determine the four-momentum of \bsig\
precisely. In order to suppress backgrounds arising from the 
dominant \B decays to charmed mesons, we require the light meson's 
momentum \pstar\ in the \bsig\ rest frame to be greater than 2.34 
(2.36) GeV in the kaon (pion) case. The separation of \Kp\ from \pip\ 
candidates is based on the Cherenkov angle measured in the Detector 
of Internally Reflected Cherenkov light. 

The \breco\ is reconstructed in the decays $B \ra D^{(*)}Y^{\pm}$,
where $Y^{\pm}$ is a combination of hadrons containing one, three,
or five charged kaons or pions, up to two neutral pions, and at most
two $\KS \ra \pip \pim$. We reconstruct $D^{* -}\rightarrow \Dzb\pi^{-}$;
$\Dstarzb \rightarrow \Dzb\pi^{0}$; $\Dzb\rightarrow K^{+}\pi^{-}$, 
$K^{+}\pi^{-}\pi^{0}$, $K^{+}\pi^{-}\pi^{-}\pi^{+}$, $\KS\pi^{+}\pi^{-}$; and
$D^-\rightarrow K^{+}\pi^{-}\pi^{-}$, $K^{+}\pi^{-}\pi^{-}\pi^{0}$, 
$\KS\pi^{-}$, $\KS\pi^{-}\pi^{0}$, $\KS\pi^{-}\pi^{-}\pi^{+}$.
We define the purity of a particular mode as $S/(S+B)$, where
$S$ ($B$) denotes the number of signal (background) events; we
use only the 186 \breco\ final states with purity, measured in
data control samples, greater than 0.2. When more than one \breco\
candidate is found in an event, the one with the highest purity
is retained; the overall purity of our selected sample is approximately 0.45.

Two kinematic variables characterize correctly reconstructed $B$ 
candidates: the energy-substituted mass $\mes\equiv\sqrt{s/4-\pvec_B^2}$
and the energy difference $\DE \equiv E_B-\sqrt{s}/2$, where 
$(E_B,\pvec_B)$ is the \B-meson four-momentum in the \FourS\ rest 
frame. For the \breco\ candidate, we select events with $5.2500 < 
\mes < 5.2893$ GeV and we apply a mode-dependent cut on \DE. Additional 
background rejection is provided by the angle $\theta_T$, defined as 
the angle between the thrust axis of the \breco\ candidate decay 
products and the rest of the event. For continuum events $|\cos \theta_T|$ 
peaks sharply at 1, while \BB\ events exhibit a uniform distribution. 
We select events with $|\cos \theta_T| < 0.9$. 

Finally, we combine into a Fisher discriminant \xf\ four variables
sensitive to the event shape and the production dynamics: the polar 
angles with respect to the beam axis in the \FourS\ frame of the 
\breco\ candidate momentum and of the \breco\ thrust axis, and the 
zeroth and second angular moments $L_{0,2}$ of the energy flow. 
The moments are defined by $L_j = \sum_i p_i\times\left|\cos\theta_i\right|^j,$ 
where $i$ labels a charged or neutral candidate not originating from
the decay of the \breco, $\theta_i$ is the angle with respect to the 
\breco\ thrust axis, and $p_i$ is its momentum.
 
\begin{figure}[!htbp]
\begin{center}
  \includegraphics[width=0.9\linewidth]{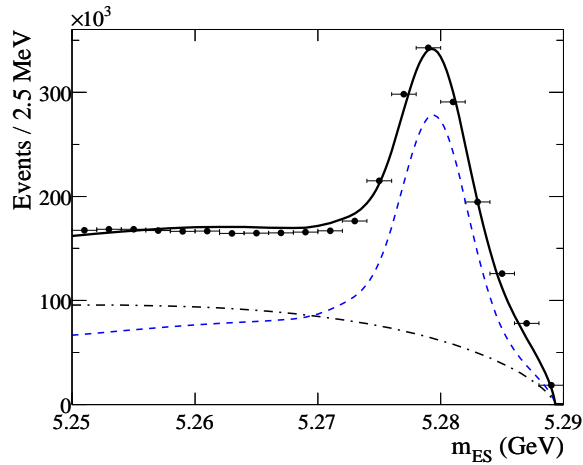}
  \caption{Projection of the \mes\ variable for the
    \breco\ sample; the dashed line represents the \BB\ component, 
    the dot-dashed is the continuum background, and the solid line
  is the sum of the two components.}
  \label{fig:proj_mes}
\end{center}
\end{figure}

The branching fractions we are measuring are normalized to the
number of fully reconstructed \BB\ events present in our sample.
We determine the \BB\ yield (over the \qqbar\ continuum background)
through a maximum likelihood fit to the variables \mes\ and \xf. 
The probability density function (PDF) of \mes\ for the \BB\ category
is the sum of two components: two Gaussian functions centered on the 
mass of the \B\ parameterize the correctly reconstructed \B\ candidates,
while an ARGUS \cite{argus} function describes the misreconstructed
$B$ decays. For the continuum we use only an ARGUS function. For the
\xf\ variable we use the sum of a bifurcated Gaussian with a Gaussian 
for both \BB\ and \qqbar. Besides the yields of the two components (\BB\ 
and \qqbar), the ARGUS exponent for the \qqbar\ component and the fraction 
of correctly reconstructed \BB\ events are free. We split the data sample 
into four subsamples characterized by different purity ranges of the 
\breco\ candidates. The ARGUS exponent and the fraction of \BB\ events 
peaking in \mes\ are allowed to take different values among these categories. 
Figure \ref{fig:proj_mes} shows the projection over the \mes\ variable of 
this fit. The \BB\ yield is $(2.0902 \pm 0.0020) \timesix$ \BB\ events. 
By repeating the fit on the subsamples with different purities and using 
different parameterizations for the PDFs, we estimate the
systematic uncertainty on the \BB\ yield to be 5\%.

We assign to \bsig\ all the charged and neutral particles not belonging 
to the \breco\ candidate and require $5.1000 < \mes(\bsig) < 5.2893$ GeV. This
loose cut suppresses background events in which a significant amount
of energy and momentum is lost. We 
suppress $b \ra c$ semileptonic decays by rejecting events where an 
electron or muon candidate is present. We also veto events in which 
a \Dz, \Dp, or \Ds candidate, with a mass within 30 MeV of the nominal 
value, is found. 

We require that a \Kp, \KS, or \pip\ candidate with $\pstar > 1.8$ GeV
be present on the signal side. The distance of closest approach for \Kp\ and
\pip\ candidates must be less than three standard deviations from the \bsig\
decay vertex. \KS\ candidates are reconstructed in the $\pip \pim$ final 
state, with requirements that the vertex probability of the two tracks be greater 
than $10^{-4}$, that the flight length be greater than three times its 
uncertainty, and that their mass satisfy $0.486 < m_{\pip\pim} < 0.510$ GeV.

We extract the signal yields from a maximum likelihood fit to the
three variables \mes(\breco), \xf, and \pstar.
For the \Kp\ and \pip\ samples we also measure the direct $CP$
asymmetry $\acp \equiv (\Gamma^- - \Gamma^+)/(\Gamma^- + \Gamma^+)$, 
where the superscript to the decay width $\Gamma$ refers to the 
charge of the light hadron. Our fits have three components:
signal, $b \ra c$ background, and continuum background. 
For each of these categories $j$ we define probability density functions 
$\calP_j(x)$ for the variable $x$, with the resulting likelihood:

\begin{eqnarray}
\calP_j  &=& \calP_j(\mes) \calP_j(\xf) \calP_j(p^{\star})  \, , \\
{\cal L} &=& \frac{e^{-\sum_j Y_j}}{N!} \prod_{i=1}^N \sum_j Y_j \calP_j^i \, ,
\end{eqnarray}
where $\calP_j^i$ is $\calP_j$ evaluated for event $i$, $Y_j$ 
is the yield for category $j$, and $N$ is the number of
events entering the fit. We assume the PDFs for each variable to be 
uncorrelated in the signal and $b \ra c$ components (a correlation in
the continuum component is handled as discussed below). We check this 
assumption by means of Monte Carlo (MC) experiments
\cite{geant}, in which signal and $b \ra c$ events are taken from 
fully simulated event samples and the continuum background is generated 
from the PDFs. In the extraction of the signal yields, we correct for 
the small biases we observe in these ensembles. The PDFs are 
extracted by fitting MC samples, where the charmless decays are 
separated from $b \ra c$ background using information at the
generator level. 

\begin{figure*}[!htbp]
\begin{center}
  \includegraphics[width=1.0\linewidth]{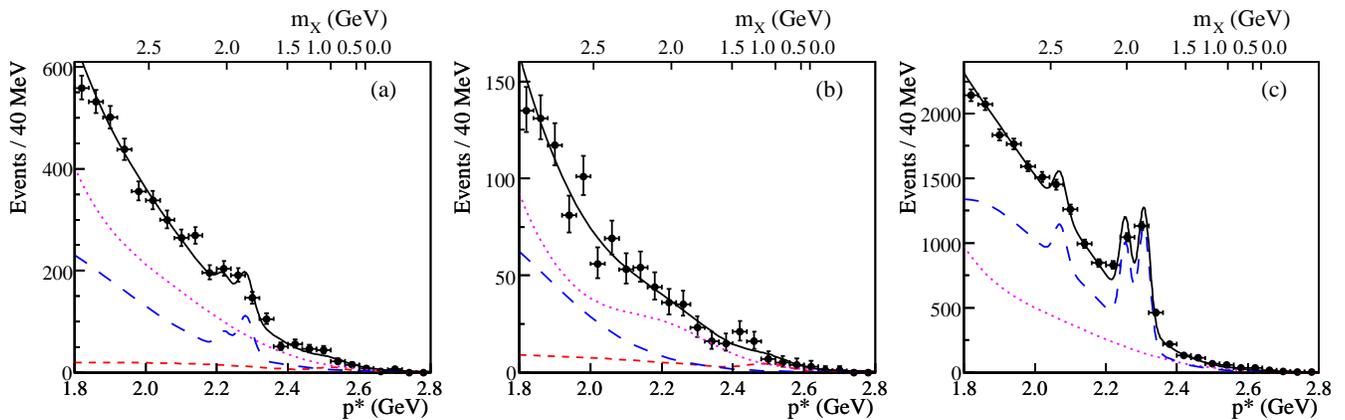}
  \caption{Projection plots for the whole \pstar\ range for the (a)  
\Kp, (b) \KS, and (c) \pip\ samples. The solid curves are the total fit 
functions, the red dashed lines are the signal components (which are
kept fixed at this stage), the blue long dashed lines are the $b \ra c$ 
background and the magenta dotted lines are \qqbar. The scale on the 
upper border of the plots indicates the mass of the system recoiling 
against the light hadron.}
  \label{fig:pstar_180}
\end{center}
\end{figure*}

Signal and $b \ra c$ events share the same PDFs for
the \mes\ and \xf\ variables which are only effective to separate
\BB\ events from the continuum; the fit distinguishes between
charmed and charmless $B$ decays by exploiting the differences in
the \pstar\ distributions. The \pstar\ distribution is parameterized 
by the sum of a Gaussian with an ARGUS component for the 
signal, by the sum of an exponential and a Gaussian for the \qqbar
component, and by the sum of three, one, or five Gaussians for the 
$b \ra c$ background in the \Kp, \KS\, and \pip\ samples, respectively.
The latter parameterize the broad component(s) of the $b \ra c$ background
and the peaking components corresponding to the $B \ra D^{(*,**)}h$,
($h = \Kp, \KS$ or \pip) decays, all of which are evident in the \pip\ 
sample (see Fig. \ref{fig:pstar_180}). Similarly, the Gaussian component 
of the signal \pstar\ PDF accounts for the dominant two-body decays (mainly
\etapK), while the broad component describes the sum of the other
contributions.
The splitting of the data into subsamples based on the purity and
the charge of the \breco\ candidates allows differences in the background
distributions to be accommodated in the fit by allowing the parameters
most sensitive to these variations to take different values in each
subsample.

The fit is performed through an iterative procedure. In the first step we 
fix the signal yield to the predictions of the MC and fit the
$\pstar > 1.8$ GeV sample, leaving free to vary the most important 
parameters of the background such as the normalization of the
peaking components in the $b \ra c$ background, the width of the
broad components, and the exponent of the ARGUS function.
This step is aimed at determining the shape and the normalization of 
the $b \ra c$ background; the projection plots for this step of the 
fit are presented in Fig.~\ref{fig:pstar_180}.

\begin{table*}[!bth]
\caption{Summary of the fit results to the high \pstar\
range. The $b \ra c$ background yield is kept fixed in this
fit; the quoted uncertainty represents the amount by which this
quantity is varied for the evaluation of systematic uncertainties. 
The first error in the branching fractions and in the direct charge
asymmetries is the statistical one, while the second is systematic
(the significance includes only the additive part of the latter). 
The upper limits (U.L.) on the partial 
branching fractions are taken at the 90\% confidence level. For the 
\pip\ sample, the results of the yields refer to the $\pstar > 2.36$ 
GeV range, whereas the branching fraction has been extrapolated to 
$\pstar > 2.34$ GeV.}
\label{tab:results}
\begin{tabular}{l|ccc}
\dbline
 & $B \ra \Kp X$ &  $B \ra \Kz X$ &  $B \ra \pip X$ \\
\sgline
Events to fit   & 306   & 84 & 692 \\
$b \ra c$ yield (events)  & $66 \pm 8$   & $6.5 \pm 2.6$ & $173 \pm 13$ \\
\qqbar\ yield (events)  & $194 \pm 15$   & $48 \pm 8$  & $430 \pm 22$ \\
Signal yield (events)  & $54^{+11}_{-10}$ & $32^{+7}_{-7}$ & $107^{+15}_{-14}$ \\
Fit bias (events) & $+10.9$   & $+3.5$ & $-4.3$ \\
Significance ($\sigma$) & \signifBKpX  & \signifBKsX & \signifBPiX \\
${\cal B}$ $(\timemsix)_{\pstar > 2.34 \mbox{ \scriptsize GeV}}$ & \pBRBKpX  & \pBRBKsX & \pBRBPiXn \\
${\cal B}$ U.L. $(\timemsix)_{\pstar > 2.34 \mbox{ \scriptsize GeV}}$ & \ulpBRBKpX & \ulpBRBKsX & $-$ \\
\sgline
\:$\acp$ & \AcpBKpX  & $-$ & \AcpBPiX \\
\dbline
\end{tabular}
\end{table*}

\begin{figure*}[!htbp]
\begin{center}
  \includegraphics[width=1.0\linewidth]{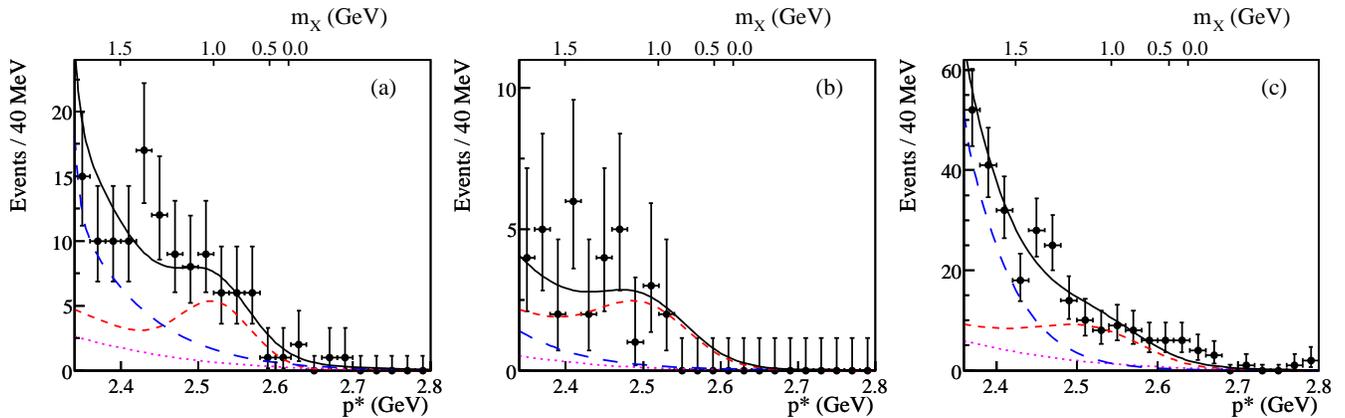}
  \caption{Projection plots for $\pstar > 2.34 (2.36)$ GeV for the (a)
\Kp, (b) \KS, and (c) \pip\ samples. The solid curves are the total fit 
function, the red dashed lines are the signal component, the blue long dashed 
are the $b \ra c$ background and the magenta dotted are \qqbar. In
order to enhance the signal component we apply cuts on the likelihood
(computed excluding the \pstar\ variable) which retain 82 \textendash 88\% of 
signal events while suppressing most of the \qqbar\ background.
The scale on the upper border of the plots indicates the mass of
the system recoiling against the light hadron.}
  \label{fig:pstar_234}
\end{center}
\end{figure*}

In the next step, we use the results obtained in the previous fit
to extrapolate the predicted $b \ra c$ background into the high \pstar\ 
region ($\pstar > 2.34$ GeV for \Kp\ and \KS, $\pstar> 2.36$ GeV for \pip).
We fit these subsamples, varying only the yields of the signal 
and \qqbar\ background components and the charge asymmetries, while
the shapes are those determined in the previous step (see Fig. 
\ref{fig:pstar_234}). An exception
occurs for the \xf\ variable in the \qqbar\ background, which is
correlated with \pstar; thus, fixing its shape to that
determined in the whole \pstar\ range would lead to a bias. 
In this case we parameterize the \xf\ distribution with two Gaussians,
determine its parameters from the MC in the high \pstar\ range,
and leave the mean of the core Gaussian free to vary in the fit.
Using the \pstar\ cut efficiency derived from the
MC, we then recalculate the number of signal events in the whole
\pstar\ range and repeat the fitting procedure from the beginning.

We find that this procedure converges after at most six cycles
and that the result does not depend on the initial values we choose
for the signal yield. We use the results of the final fit to the 
high \pstar\ range to derive the partial branching fractions
and the direct $CP$-asymmetries (for the \Kp\ and \pip\ samples).
The branching fractions are computed using the efficiencies for
reconstructing signal events in the high \pstar\ region derived 
from the simulation. In order to avoid the systematic uncertainty 
related to the \breco\ reconstruction efficiency, the calculation
is done taking for the normalization the number of \BB\ events present
in our sample.
To make the comparison with the kaon samples easier, we 
extrapolate the branching fraction of \BPiX\ to the $\pstar > 2.34$ 
GeV range (we assume the systematic error associated with this extrapolation 
to be negligible).  The results are collected in Table \ref{tab:results}.

The whole fit procedure is tested on a data sample enriched in
$b \ra c$ background, selected by reversing the vetoes on the \Dz, 
\Dp, or \Ds\ candidates associated with the \bsig. The results agree 
within statistical uncertainties with the expectations of very
small signal yields. We also verify that our model for the 
continuum background is in very good agreement with the data taken
away from the \FourS\ resonance.

Systematic uncertainties arise from the imperfect knowledge of
the number of \breco\ candidates (5\%), from the uncertainties
on the reconstruction efficiencies for charged particles (0.5\%), 
\KS\ candidates (2.1\%), and other neutral particles 
(0.9\textendash1.2\%, depending on the final state), from
the $K/\pi$ separation (2.4\%), and from the statistics of the MC sample 
which we use to compute the efficiency in reconstructing signal 
events (6.8\textendash14.5\%). The above uncertainties are multiplicative
and do not affect the significance of the measured branching fractions,
contrary to the following additive contributions: the uncertainty on the 
PDFs of the signal component is estimated by leaving each parameter kept 
fixed in the nominal fit free to vary (3.6\textendash8.5 events). The 
uncertainty in the $b \ra c$ background is computed by varying its yield 
by the sum in quadrature of its Poisson uncertainty and the uncertainty 
in the extrapolation to the high \pstar\ region, taking into account the 
uncertainty on the knowledge of the signal PDF. The resulting systematic 
error is 2.8\textendash10.3 events. The systematic error arising from the 
correction for the fit bias is taken as the sum in quadrature of half 
the correction itself and the statistical uncertainty on the correction 
(3.6\textendash7.9 events).

The systematic uncertainties for the direct $CP$-asymmetries include
the uncertainty in detector related charge asymmetries, which mainly
affect the kaons (2\%), different reconstruction efficiencies for
\B\ and \Bb\ candidates in the tag sample (2.5\%), and effects due
to mistagging (3\%).

Our results for the partial branching fractions and \acp\ are given with
statistical and systematic errors in Table \ref{tab:results}.  The
central values for the branching fractions are in agreement with our
estimates \cite{sumBF} of the
sums of the known exclusive branching fractions of charmless two- and
three-body $B$-decays. On the other hand, predictions based on SCET 
\cite{chay} underestimate the measurements, both those 
of the inclusive branching fractions presented here and those obtained 
by summing exclusive modes, even after adjusting for the branching 
fractions of the $B \ra \eta^{(\prime)} X$ modes, which are acknowledged 
to be problematic for SCET.
This fact is interpreted by the authors of Ref. \cite{chay} as an 
indication of the need to introduce substantial nonperturbative charming 
penguin contributions or large higher-order corrections.

In conclusion we have measured the inclusive partial branching fractions
for \BKpX, \BKzX, and \BPiX\ in the region where the momentum of
the candidate hadron is greater than 2.34 GeV. 
The statistical significance, computed as the difference between 
the value of $-2\ln{\cal L}$ for the zero signal hypothesis and the 
value at its minimum, exceeds five standard deviations in each case; 
however, comparable systematic uncertainties lower the significance to
the values quoted in the Table, and we quote 90\% confidence level
upper limit (taken as the value below which lies 90\% of the total 
of the likelihood integral, in the region where the branching fraction 
is positive) for the \Kp\ and \Kz\ modes.   
All results are in agreement with the standard model predictions,
and exclude large enhancements due to sources of new physics.
We do not find any significant direct $CP$-asymmetry in the \Kp\ 
and \pip\ samples.

\par\input pubboard/acknow_PRL


\renewcommand{\baselinestretch}{1}

\end{document}

%% file: Q2BDefn/Definitions.tex
\RequirePackage{xspace}

\newcommand{\calP}{\ensuremath{{\cal P}}}

\newcommand{\pvec}{{\bf p}}

\newcommand{\acp}{\ensuremath{\calA_{\rm ch}}}

\newcommand{\timesix}{\ensuremath{\times10^{6}}}
\newcommand{\timemsix}{\ensuremath{\times10^{-6}}}

\newcommand{\DE}{\ensuremath{\Delta E}}

\newcommand{\xf}{\ensuremath{{\cal F}}}

\newcommand\etal{{\it et al.}}

\newcommand{\bfig}{\begin{figure}[htbpc!]}
\newcommand{\efig}{\end{figure}}
\newcommand\bef{\begin{figure}}
\newcommand\edf{\end{figure}}
\newcommand\dbline{\noalign{\vskip 0.10truecm\hrule}\noalign{\vskip 2pt}\noalign{\hrule\vskip 0.10truecm}}

\newcommand\sgline{\noalign{\vskip 0.10truecm\hrule\vskip 0.10truecm}}
\newcommand\beq{\begin{equation}}
\newcommand\eeq{\end{equation}}
\newcommand\bear{\begin{array}}
\newcommand\enar{\end{array}}
\newcommand\beqa{\begin{eqnarray}}
\newcommand\eeqa{\end{eqnarray}}
\newcommand\ben{\begin{enumerate}}
\newcommand\een{\end{enumerate}}

\newcommand{\fetapK}{\ensuremath{\etapr K}}
\newcommand{\etapK}{\ensuremath{\B\ra\fetapK}}

%% file: pubboard/authors_oct2010_bad2317.tex
%
\author{P.~del~Amo~Sanchez}
\author{J.~P.~Lees}
\author{V.~Poireau}
\author{E.~Prencipe}
\author{V.~Tisserand}
\affiliation{Laboratoire d'Annecy-le-Vieux de Physique des Particules (LAPP), Universit\'e de Savoie, CNRS/IN2P3,  F-74941 Annecy-Le-Vieux, France}
\author{J.~Garra~Tico}
\author{E.~Grauges}
\affiliation{Universitat de Barcelona, Facultat de Fisica, Departament ECM, E-08028 Barcelona, Spain }
\author{M.~Martinelli$^{ab}$}
\author{D.~A.~Milanes}
\author{A.~Palano$^{ab}$ }
\author{M.~Pappagallo$^{ab}$ }
\affiliation{INFN Sezione di Bari$^{a}$; Dipartimento di Fisica, Universit\`a di Bari$^{b}$, I-70126 Bari, Italy }
% ...
\collaboration{The \babar\ Collaboration}
\noaffiliation

%% file: pubboard/acknow_PRL.tex
We are grateful for the excellent luminosity and machine conditions
provided by our \pep2\ colleagues, 
and for the substantial dedicated effort from
the computing organizations that support \babar.
The collaborating institutions wish to thank 
SLAC for its support and kind hospitality. 
This work is supported by
DOE
and NSF (USA),
NSERC (Canada),
CEA and
CNRS-IN2P3
(France),
BMBF and DFG
(Germany),
INFN (Italy),
FOM (The Netherlands),
NFR (Norway),
MES (Russia),
MICIIN (Spain),
STFC (United Kingdom). 
Individuals have received support from the
Marie Curie EIF (European Union),
the A.~P.~Sloan Foundation (USA)
and the Binational Science Foundation (USA-Israel).

%% file: BAD-2317.bbl
\begin{thebibliography}{99}
\bibitem{CC}
Unless otherwise stated, charge conjugate reactions are implied
throughout the paper.
\end{thebibliography}
